\documentclass[final, runningheads]{llncs}
\usepackage[T1]{fontenc}
\usepackage{graphicx}
\usepackage{amsmath}
\newcommand{\itadata}{\footnotesize \textsl{ITADATA2024: The 3$^{\text{rd}}$ Italian Conference on Big Data and Data Science}}
\usepackage{fancyhdr}
\fancypagestyle{empty}{\fancyhf{}\fancyhead[C]{\itadata}
  \fancyhead[R]{\includegraphics[width=.05\textwidth]{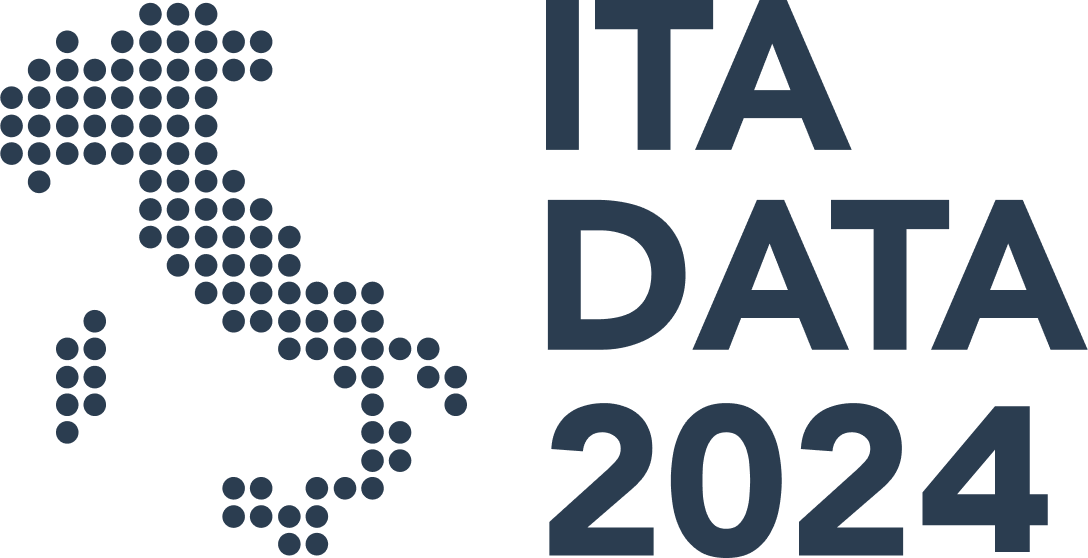}}}

\usepackage{amssymb}
\usepackage{algorithm}\usepackage{varwidth}\usepackage{graphicx}
\usepackage{multirow}
\usepackage{algorithm}
\usepackage{comment}
\usepackage{xcolor}
\usepackage{subcaption}
\usepackage{makecell}
\usepackage{listings}
\usepackage{pgfplots}
\pgfplotsset{compat=1.18}
\usepackage{tikz}
\usepackage[colorinlistoftodos]{todonotes}

\newcommand{\cdtodoin}[1]{\todo[inline,caption=0,color=red!40]{\textbf{CD:} #1}{}}
\newcommand{\amtodoin}[1]{\todo[inline,caption=0,color=green!40]{\textbf{AM:} #1}{}}

\usepackage{hyperref}
\usepackage{algorithmicx}
\usepackage{comment}
\usepackage{array}
\usepackage{booktabs}
\usepackage[noend]{algpseudocode}

\makeatletter
\begin{document}
\title{A metadata model for profiling multidimensional sources in data ecosystems}

\author{Claudia Diamantini\inst{1} \and Alessandro Mele\inst{1} \and Domenico Potena\inst{1}\and \\ Cristina Rossetti\inst{1}\inst{2} \and Emanuele Storti\inst{1}}\authorrunning{C. Diamantini et al.}
\institute{DII, Università Politecnica delle Marche, Via Brecce Bianche, Ancona 60131, Italy
\and DAUIN, Politecnico di Torino, 
Corso Duca degli Abruzzi 24, Torino, 10129,
Italy
\email{\{c.diamantini,d.potena,e.storti\}@univpm.it,\\ \{a.mele,c.rossetti\}@pm.univpm.it}
}
\maketitle              
\begin{abstract}
The Big Data landscape poses challenges in managing diverse data formats, requiring efficient storage and processing for high-quality analysis. Effective metadata management is crucial for organizing, accessing, and reusing data within these data ecosystems.
Existing metadata vocabularies and standard, however, do not adequately accommodate aggregated or summary data. This paper introduces a metadata model to support semantic annotation and profiling of multidimensional data. Defined as an RDF vocabulary, the model provides a flexible and extensible graph representation for metadata at source and attribute levels, aligning dimensions and measures to a reference Knowledge Graph and summarizing value distributions in profiles. 
An evaluation of the execution time for profile generation is also proposed, across data sources with different cardinalities.
\end{abstract}

\section{Introduction}
\label{sec:introduction}
Big Data world poses several challenges due to the management of large amount of heterogeneous data with different formats. Data need to be efficiently stored and processed in order to ensure the quality of data analysis. 
In data ecosystems, both centralized like Data Lakes, and decentralized Data Spaces, metadata management is crucial for several reasons, as it helps ensuring that data are properly organized, easily findable, accessible, interoperable and reusable \cite{wilkinson2016fair}. 
Existing metadata vocabularies and standards are however not tailored to  aggregated or summary data, which are characterized by a multidimensional nature and a specific structure \cite{hoseini2024survey}.
Aggregated data, which consolidate information from multiple sources into a unified format, are essential for monitoring, management, and planning purposes across various sectors. 
This type of data serves as a support to  decision-making processes in both public and private sectors, offering insights that are crucial for strategic operations and policy formulation.
Examples include open data by public bodies, e.g., to  monitor economic trends or the effectiveness of governmental policies and initiatives.

In this paper, we introduce a metadata model to support semantic annotation and profiling of summary data. The model is defined as an RDF vocabulary, to uniformly represent different sources through a flexible and extensible graph representation that can be easily queried, shared, and integrated with other metadata sources.
 The model allows to specify metadata at source and attribute level, including the alignment of dimensions and measure to concepts of a reference Knowledge Graph. Furthermore, defined metadata include statistical information for attributes and a summarization of value distributions in profiles. 
 These features enable data discovery and exploration functionalities such as dataset discovery based on specific data characteristics or data quality assessment, that are not supported by other models.
 The work is complemented by an evaluation of execution time for profile generation with data sources of different cardinalities.

The rest of this paper is organised as follows: in Section \ref{sec:related_work} we discuss relevant related work. The metadata model is introduced in Section \ref{sec:methodology}, while an evaluation is proposed in Section \ref{sec:experiments}. 
Finally, Section \ref{sec:conclusion} concludes the paper and outlines possible future developments.

\section{Related work}
\label{sec:related_work}
\begin{comment}
A Data Lake, according to Khine et al.\cite{khine2018data}, can easily turns into a Data swamp, a repository that contains data difficult to retrieve, and inevitably analyse; to avoid this, a proper governance, including an efficient metadata management system\cite{miloslavskaya2016big} is required.
Since the data sources are ingested with no pre-elaboration, it's important to collect metadata in order to make data Findable, Accessible, Inter-operable, and Reusable (FAIR)\cite{wilkinson2016fair}.
\end{comment}
Metadata play a crucial role in organizing and describing vast amounts of data within data ecosystems, enabling users to locate, understand, exchange and share datasets effectively. 
To this aim, effective data governance requires a proper metadata model adhering to the FAIR (Findable, Accessible, Interoperable, and Reusable) principles \cite{wilkinson2016fair}.

In the context of Data Lake literature, metadata are primarily categorized in two types: functional and structural \cite{sawadogo2021data}.
Within the former category, metadata can be further classified as (i) technical (i.e., data format, structure or schema), (ii) operational (i.e., data location, file size, number of record) and (iii) business (i.e., field names and explanations) \cite{oram2015managing}.
Diamantini et al.\cite{diamantini2018new} show that these categories are not entirely distinct and can overlap.
In the latter category, metadata are associated with the concept of object \cite{sawadogo2021data}, a generalisation of a data source that can be structured (e.g., SQL tables), semi-structured (e.g., XML, JSON), or unstructured  (e.g., images, audio or video).
Sawadogo et al.\cite{sawadogo2019metadata} categorised metadata in (i) intra-object, related to general properties of an object (i.e., file name, size, and location, semantic annotation), (ii) inter-object, for representing relations between objects (e.g., grouping through metadata or similarity links using affinity and joinability measures) and (iii) global metadata (e.g., semantic resources, indexes and logs tracking user activities in the data lake). We observe that 
such typologies include most of the elements identified by Oram et al.\cite{oram2015managing}. Specifically, business metadata are comparable to semantic metadata, operational metadata can be considered as logs, technical metadata are equivalent to pre-visualisation metadata, and structural metadata can be seen as an extension of functional metadata.

Metadata are also core components of other data sharing ecosystems like Data Space architectures. In fact, Data Spaces add decentralized governance to Data Lakes, facilitating the sharing and exchange of data across different organizations without the need of a single, centralized repository. This makes FAIR principles even more important and challenging. Data space architectures typically introduce 
components for registration, publication, maintenance, and query of metadata (e.g., 
IDS-RAM metadata broker \cite{IDS-RAM4}, EOSC's Metadata Framework \cite{EOSC}, GAIA-X Federated Catalogue \cite{GAIA-X}). Given the decentralized,  market-based nature of Data Spaces, more emphasis is given to the description of identities, providers, and services than in more centralized Data Lake scenarios like ours. On the other hand, interoperability is enabled by adopting standard metadata formats and vocabularies, following Semantic Web principles, which characterize our proposal as well.

Several vocabularies and ontologies have been proposed in the literature to describe (statistical) data sources \cite{hoseini2024survey}. Among them,
``Vocabulary of Interlinked Datasets'' (VoID)\footnote{https://www.w3.org/TR/void/} is an RDF Schema vocabulary for expressing high-level metadata about RDF datasets, such as the vocabularies used in the dataset and the size of the dataset. 
Similarly, Dublin Core Metadata Initiative (DCMI) Terms\footnote{https://www.dublincore.org/specifications/dublin-core/dcmi-terms/\#} aims at interoperable metadata descriptions for resources in the Web, also including basic data provenance metadata. 
RDF Data Cube\footnote{https://www.w3.org/TR/vocab-data-cube/} is an RDF Schema vocabulary based on the SDMX standard designed for representing statistics in a multidimensional attribute space, while  \texttt{DCAT}\footnote{https://www.w3.org/TR/vocab-dcat-3/} is an RDF vocabulary designed to facilitate interoperability between data catalogues published on the Web. For each source within the catalogue, some basics metadata are provided, i.e., resource URL, title, file type and dimension.
For what concerns tools for statistics generation, LODstat \cite{auer2012lodstats} provides schema and data statistics for RDF data sources, i.e., number of triples, triples containing blank nodes, or literal values with the respective data types.
PROLOD++ \cite{abedjan2014profiling} is a tool for profiling and mining RDF data sources. Profiles provide statistics such as frequency counts and distributions related to subjects, predicates and objects.
Similarly, RDFStats \cite{langegger2009rdfstats} is a generator for statistics of RDF data sources, which provides histograms of classes, properties and value types.
We emphasise that the mentioned  vocabularies and tools do not provide support for the type of metadata we are interested in, since they are not tailored to the description of the peculiar characteristics of multidimensional data sources, namely  measures and dimensions.

\section{Metadata model}
\label{sec:methodology}

This section discusses a metadata model for multidimensional data sources in a data ecosystem, providing a structured and interoperable framework to describe sources and their attributes.
Hereby, we represent a source $\cal S$ as a set $\{a_1,\ldots,a_n\}$ of attributes. By referring to the multidimensional model, attributes can be categorized as:
\begin{itemize}
\item dimensions, i.e. categorical attributes providing different perspectives for the analysis of data (e.g., Time, Location, Product), which are are organized in a hierarchy of levels;
\item measures, i.e. quantitative attributes that can be aggregated and analyzed (e.g., Sales Amount, Quantity Sold);
\item descriptive attributes, i.e. additional context-providing attributes (e.g., creation date, creator).
\end{itemize}

For the semantic annotation of multidimensional attributes of a data source, we assume the availability of a Knowledge Graph providing a representation of the domain knowledge in terms of definitions of measures, dimensions organized in hierarchies of levels, and members.
The Graph relies on the KPIOnto ontology, an OWL2-RL ontology aimed to provide classes and properties to model measures (named ``indicators'' in the ontology), dimensions and further information \cite{diamantini2024analytic}.
The metadata model itself employs a graph-based representation using the Resource Description Framework (RDF), allowing for the uniform representation of heterogeneous sources. This accommodates both structured data sources (such as SQL databases) and semi-structured data sources (such as JSON or XML files) in a consistent manner. This uniformity simplifies the integration and querying of diverse datasets, thus enhancing overall interoperability.
As such, both a data source and its attributes are uniquely identified by  Uniform Resource Identifiers (URIs).
We refer hereby to classes and properties from the imported vocabularies through the prefixes and corresponding namespaces reported in Table \ref{table:namespaces}. We refer to prefix \texttt{dl} for the metadata model, and to prefix \texttt{kg} for local data.

\begin{comment}Concepts and relationships are represented using different vocabularies: \texttt{rdfs}\footnote{https://www.w3.org/TR/rdf12-schema/}, \texttt{xsd}\footnote{https://www.w3.org/TR/xmlschema11-1/\#intro} and {\texttt{void}\footnote{https://www.w3.org/TR/void/}}, defined by W3C\footnote{https://www.w3.org}, \texttt{dcterms}\footnote{https://www.dublincore.org/specifications/dublin-core/dcmi-terms/} defined by Dublin Core\footnote{https://www.dublincore.org} and  \texttt{KPIOnto}\footnote{https://kdmg.dii.univpm.it/kpionto/specification/}. We also introduce a custom vocabulary \texttt{dl} to describe the properties which we will discuss in the following.
\end{comment}
\begin{table}[tb]
\begin{center}
\begin{tabular}{|l|l|l|}
\hline
\textbf{Vocabulary} & \textbf{Prefix} & \textbf{Namespace} \\
\hline
XML Schema & \texttt{xsd}  & \texttt{http://www.w3.org/2001/XMLSchema\#} \\
\hline
Dublin Core Terms & \texttt{dcterms}  & \texttt{http://purl.org/dc/terms/} \\
\hline
Vocabulary of Interlinked Datasets & \texttt{void}  & \texttt{http://rdfs.org/ns/void\#} \\
\hline
KPIOnto & \texttt{kpi}  & \texttt{http://w3id.org/kpionto/} \\
\hline
Data source metadata model & \texttt{dl} & \texttt{http://kdmg.dii.univpm.it/dl/} \\
\hline
\end{tabular}
\caption{List of vocabularies with their respective prefixes and namespaces.}
\label{table:namespaces}
\end{center}
\end{table}

In the following, we discuss two types of metadata: those related to the source, including general statistics and structural information, and those related to attributes, specifying possible mappings to the Knowledge Graph, along with their profile.

\subsubsection{Data source.}
\begin{figure}[tbp]
\centering
\begin{subfigure}[b]{0.3\textwidth}
\includegraphics[width=\textwidth]{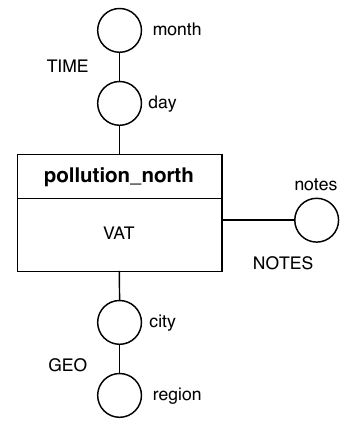}
\caption{}
\label{fig:fact_table}
\end{subfigure}
\begin{subfigure}[b]{0.6\textwidth}
\includegraphics[width=\textwidth]{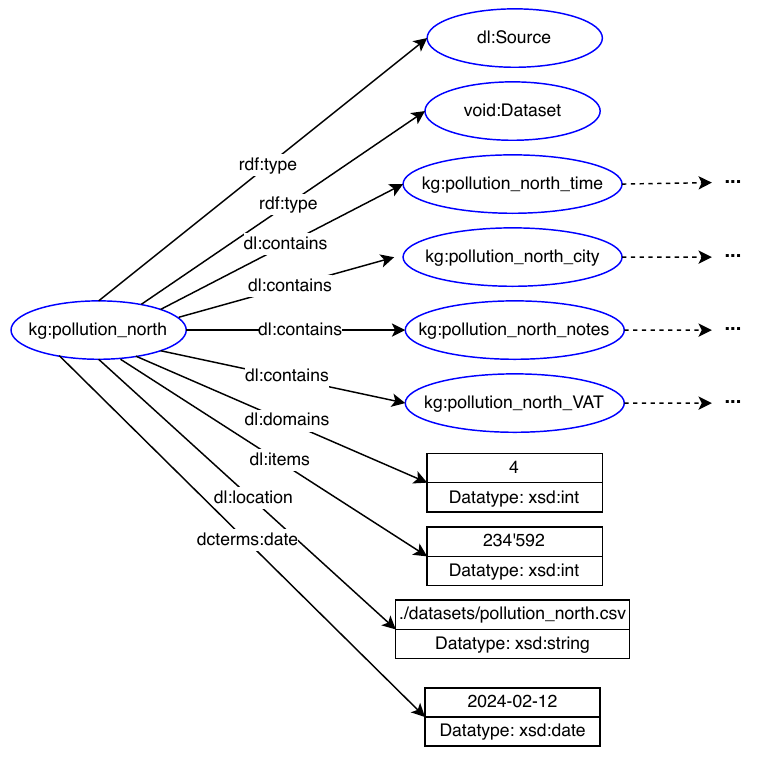}
\caption{}
\label{fig:source_profile}
\end{subfigure}

\caption{(a) Conceptual representation of a data source according to the Dimensional Fact Model and (b) the corresponding data source metadata. Dashed lines link to attributes  metadata.}
\label{fig:E}
\end{figure}

A data source is represented as an instance of class \texttt{dl:Source}.
The class extends the class \texttt{void:Dataset} from the VoID vocabulary, to provide enhanced integration capabilities.
Metadata are represented through a set of properties and corresponding values that are  linked to a source node. 
They allow to represent descriptive information included in the DCMI metadata  such as the \texttt{title} of the source, its \texttt{description}, the file \texttt{format}, a number of \texttt{subject}s (typically linking to DBpedia resources), the \texttt{creator}, the \texttt{publisher} and possible \texttt{contributor}s, the creation \texttt{date}, the \texttt{licence}.
Basic statistics such as the number of dimensions are represented through the property \texttt{dl:domains}, while the cardinality through the property \texttt{dl:items}.
Further metadata include the file path of the data source (\texttt{dl:location}), provenance and lineage information, and structural metadata. 
As for structural metadata, an attribute is defined as an instance of the \texttt{dl:Domain} class, to which a source is connected through the property \texttt{dl:contains}.
Table \ref{table:source_metadata} shows a subset of the  properties with \texttt{domain} and \texttt{range}. To give an example, Figure \ref{fig:fact_table} shows the conceptual representation of a simple source according to the Dimensional Fact Model \cite{DFM98}. The source provides data about some northern Italian regions, specifically the number of vehicles passing on city roads ('Vehicles At Time' - 
VAT indicator) which is aggregated along the time and geographic dimensions, each organized in a simple hierarchy of two levels (day and month for the former, city and region for the latter). Note the use of a degenerate dimension to represent the descriptive attribute notes storing textual descriptions about relevant events occurred (e.g., a blackout).  Figure \ref{fig:source_profile} shows the source metadata for the example: properties link a node of type \texttt{dl:Source} labeled with the source name to (i) literals for basic properties (i.e., items, location, date) and (ii)  elements of the class \texttt{dl:Domain} for each attribute in the source. 
\begin{table}[ht]
\begin{center}
\resizebox{.65\textwidth}{!}{
\begin{tabular}{|l|l|l|l|} 
\hline
\textbf{Item} & \textbf{Domain} & \textbf{Property} & \textbf{Range}\\
\hline
Location & dl:Source & dl:location & rdf:Literal \\
\hline
    Number of attributes & dl:Source & dl:domains & rdf:Literal \\
    \hline
    Cardinality & dl:Source & dl:items & rdf:Literal \\
    \hline
    Attributes & dl:Source & dl:contains & dl:Domain \\
\hline
\end{tabular}}
\caption{Excerpt of the metadata properties for a data source.}
\label{table:source_metadata}
\end{center}
\end{table}

In turn, each attribute is connected to its respective profile type.
Specifically, the model provides specialized metadata  for dimensional attributes on one side, and measures and descriptive attributes on the other. 

\subsubsection{Dimensional attributes.}
A dimensional attribute is an instance of \texttt{dl:Domain} and represents a dimension. A property \texttt{dl:mapTo} allows to map the attribute to a corresponding level (instance of \texttt{kpi:Level}) in the Knowledge Graph\footnote{We refer interested readers to \cite{diamantini2024analytic} for the definition of the automatic procedure for mapping discovery we rely on, based on the evaluation of approximate set containment.}.

A profile for dimensional attributes is defined as an instance of the \texttt{dl:DProfile} class, which in turn is a specialization of class \texttt{dl:Profile} and includes summary information on its value distribution as a set of \texttt{dl:DProfileElement}.
A profile element is defined for each value belonging to the attribute that can be mapped to a corresponding member of the dimension's level. 
Property \texttt{dl:toMember} associates the DProfileElement to a member (instance of class \texttt{kpi:Member}) in the Knowledge Graph, while property \texttt{dl:frequency} specifies the corresponding number of occurrences.
Finally, a special profile element is defined for all values that cannot be mapped to any member. In this case, instead of a  member, it is associated to a special value \texttt{dl:others}.

\begin{figure}[tb]
	\centering
\includegraphics[angle=0,origin=c,width=0.8\textwidth]{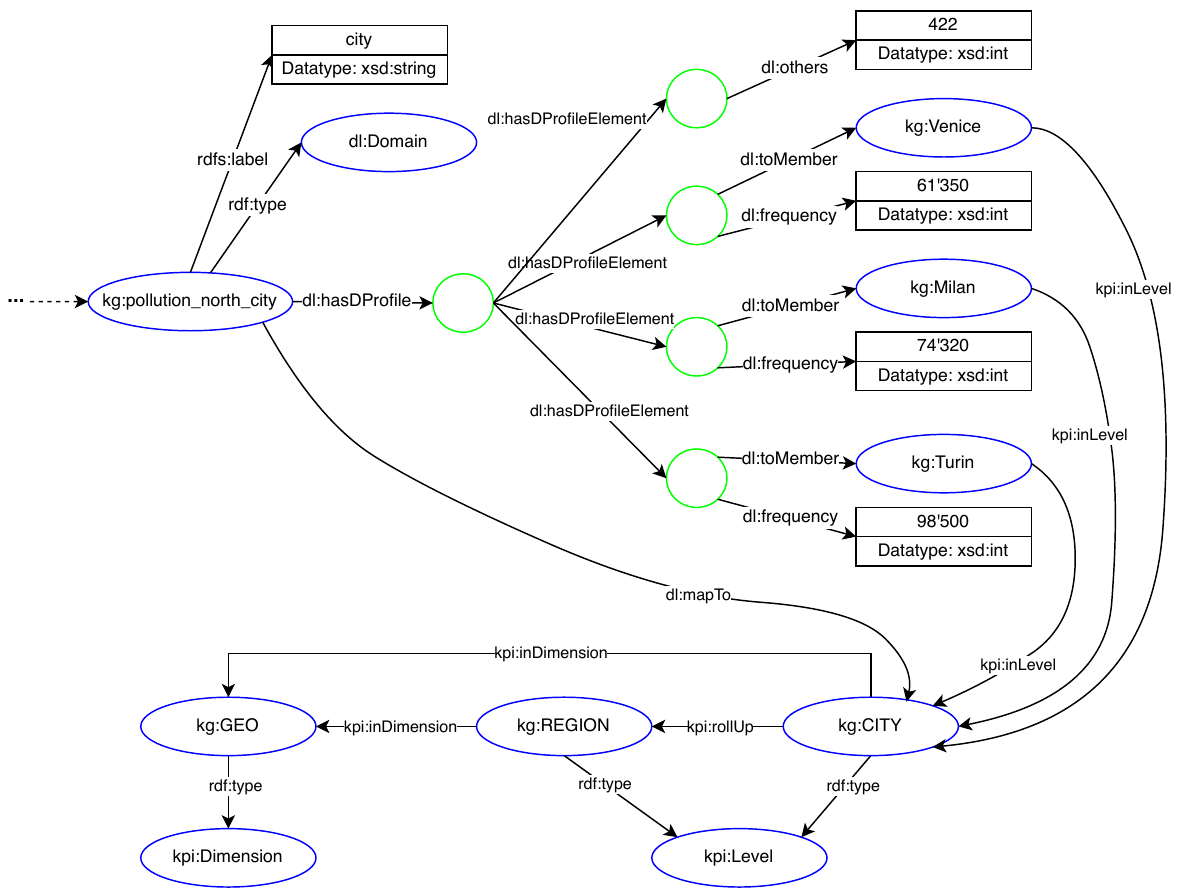}
    \caption{Example of described metadata for the dimensional attribute city with an excerpt from the Knowledge Graph involving the attribute.}
    \label{fig:level_profile}
\end{figure}

Figure \ref{fig:level_profile} shows a visual example of the described metadata for a dimensional attribute with an excerpt of relevant concepts from the Knowledge Graph, while Table \ref{table:level_profile} shows in detail, for each element, its respective properties.

\begin{table}[tb]
\begin{center}
\resizebox{\textwidth}{!}{
\begin{tabular}{|p{5.5cm}|l|l|l|} 
\hline
\textbf{Item} & \textbf{Domain} & \textbf{Property} & \textbf{Range}\\
\hline
Profile associated with the attribute & dl:Domain & dl:hasDProfile & dl:DProfile \\
\hline
Level in the Knowledge Graph with which the attribute is associated & dl:Domain & dl:mapTo & kpi:Level \\
\hline
Member of the level with which a profile element is associated & dl:DProfileElement & dl:toMember & kpi:Member \\
\hline
Number of occurrences of the member occurring in the attribute & dl:DProfileElement & dl:frequency & rdf:Literal \\
\hline
Number of elements in the attribute that are not associated with members of the level. & dl:DProfileElement & dl:others & rdf:Literal \\
\hline
\end{tabular}}
\caption{Excerpt of described properties for a level attribute.}
\label{table:level_profile}
\end{center}
\end{table}

\subsubsection{Measures and Descriptive attributes.}
  A measure or a descriptive attribute is an instance of \texttt{dl:Domain} and belongs to one of the following type categories: \emph{integer}, \emph{decimal}, \emph{datetime}, \emph{textual}, and \emph{categorical}. 
  This last identifies attributes containing a number of distinct items below a certain threshold, while a further category \emph{unrecognized} is used whenever the attribute type cannot be identified. 
  Only one category can be associated with an attribute. 
Unlike other descriptive attributes, a measure is characterized by a property \texttt{dl:mapTo} which, similarly to dimensional attributes, links it to an instance of class \texttt{kpi:Indicator} in the Knowledge Graph.
 
\begin{figure}[ht]
	\centering
\includegraphics[angle=0,origin=c,width=0.8\textwidth]{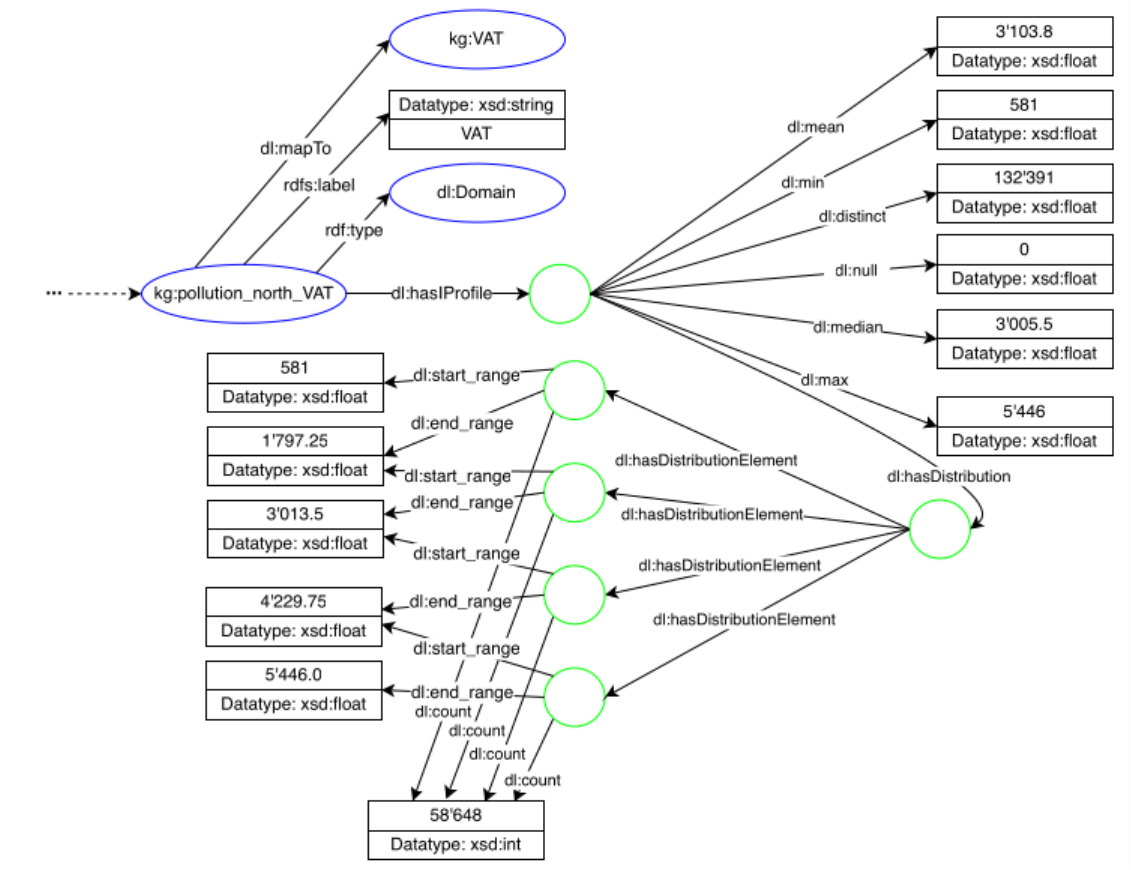}
    \caption{Example of metadata for an integer attribute representing a measure.}
    \label{fig:integer_profile}
\end{figure}

  A profile for these types of attributes is defined as an instance of class \texttt{dl:IProfile}, which in turn is a specialization of class \texttt{dl:Profile} and associated with the attribute through the property \texttt{dl:hasIProfile}. The specific set of statistics and profile elements is customized for each data type. For integer attributes, the elements comprising the profile are maximum (\texttt{dl:max}), minimum (\texttt{dl:min}), average (\texttt{dl:mean}) and median (\texttt{dl:median}), number of distinct (\texttt{dl:distinct}) and null values (\texttt{dl:null}), value distribution (class \texttt{dl:Distribu\-tion}), where for each interval (class \texttt{dl:DistributionElement}) the maximum (\texttt{dl:end\_range}) and minimum (\texttt{dl:start\_range}) values, along with their respective counts (\texttt{dl:count}), are expressed.
For categorical attributes, the elements comprising the profile are the number of null values (\texttt{dl:null}) and the details of the categories
(\texttt{dl:Categories}), where each category element (\texttt{dl:Category\-Element}) includes the category name  (\texttt{dl:category}) and its corresponding count (\texttt{dl:ca\-te\-go\-ry\-Count}).
For date attributes, described elements are the number of distinct  (\texttt{dl:distinct}) and null (\texttt{dl:null}) values, the most recent (\texttt{dl:maxDate}) and the least recent (\texttt{dl:minDate}) date, and finally the details on years (\texttt{dl:Years}), where each year element (\texttt{dl:YearElement}) includes the year itself (\texttt{dl:year}) and its corresponding count (\texttt{dl:yearCount}).
For textual attributes, described properties are the number of null values (\texttt{dl:null}), the total number of words (\texttt{dl:words}) and details on the words (\texttt{dl:Words}), where each word element (\texttt{dl:Word\-Element}) includes the word name (\texttt{dl:word}) and its corresponding count (\texttt{dl:wordCount}). For lack of space, we only illustrate the properties relating to an integer attribute in Table \ref{table:integer_profile} with a visual example in Figure \ref{fig:integer_profile}.
\begin{table}[tb]
\begin{center}
\resizebox{\textwidth}{!}{
\begin{tabular}{|>{\raggedright\arraybackslash}p{2.5cm}|l|l|l|} 
\hline
\textbf{Item} & \textbf{Domain} & \textbf{Property} & \textbf{Range}\\
\hline
Profile & dl:Domain & dl:hasIProfile & dl:IProfile \\
\hline
Maximum value & dl:IProfile\ & dl:max & rdf:Literal \\
\hline
Minimum value & dl:IProfile\ & dl:min & rdf:Literal \\
\hline
Average value & dl:IProfile\ & dl:mean & rdf:Literal \\
\hline
Median & dl:IProfile\ & dl:median & rdf:Literal \\
\hline
Number of distinct values & dl:IProfile\ & dl:distinct & rdf:Literal \\
\hline
Number of null values & dl:IProfile\ & dl:null & rdf:Literal \\
\hline
Value distribution & dl:IProfile\ & dl:hasDistribution & dl:Distribution \\ \hline
Value interval & dl:Distribution & dl:hasDistributionElement & dl:DistributionElement \\ \hline
Lower bound & dl:DistributionElement & dl:start\_range & rdf:Literal \\ \hline
Upper bound & dl:DistributionElement & dl:end\_range & rdf:Literal \\ \hline
Number of elements in the interval & dl:DistributionElement & dl:count & rdf:Literal \\
\hline
\end{tabular}}
\caption{Excerpt of described properties for an integer attribute.}
\label{table:integer_profile}
\end{center}
\end{table}
\subsubsection{Use cases.}
The metadata model can find applications in various use cases, such as (i) preliminary stages of exploratory data analysis, specifically searching for sources based on attribute properties; (ii) assessment of data quality, as shown in \cite{diamantini2023assessment}; (iii) dataset discovery, which involves identifying data sources that exhibit specific characteristics; (iv) query answering, where profile metadata can support efficient identification of sources of interest; (v) data integration, which aims to integrate data sources based on semantic similarity or attribute content; (vi) source comparison, as comparing data distributions can be crucial to determine whether to join data sources or merge them into a single entity.

\section{Experiments}
\label{sec:experiments}
In this section, we evaluate the efficiency of the metadata extraction process for (i) dimensional and (ii) measure and descriptive attributes. Experiments were run on a Mac Book Pro ARM with 8-cores CPU and 16GB of RAM. We report the average values from ten executions.

\subsection{Dimensional attributes}
\label{subsec:dimensional_attributes}
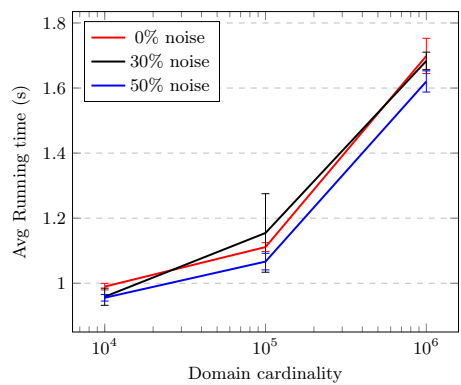
\begin{figure}[htb]
  \centering
  \resizebox{.5\columnwidth}{!}{
  \begin{tikzpicture}
    \begin{axis}[
        xlabel={Domain cardinality},
        xmode = log,
ylabel={Avg Running time (s)},
        grid style=dashed,
        legend pos= north west,
        xminorticks=true,
        ymajorgrids=true,
    ]
    \addplot[color=red, line width=1pt, mark=circle, error bars/.cd, y dir=both, y explicit]
    table[x index=1, y index=3, y error index=4, col sep=comma]{0.0_noise_level.csv};
    \addlegendentry{0\% noise}

    \addplot[color=black, line width=1pt, mark=circle, error bars/.cd, y dir=both, y explicit]
    table[x index=1, y index=3, y error index=4, col sep=comma]{0.3_noise_level.csv};
    \addlegendentry{30\% noise} 

    \addplot[color=blue, line width=1pt, mark=circle, error bars/.cd, y dir=both, y explicit]
    table[x index=1, y index=3, y error index=4, col sep=comma]{0.5_noise_level.csv};
    \addlegendentry{50\% noise} 
    \end{axis}
    \end{tikzpicture}}
  \caption{Average running time and standard deviation for profile calculation of dimensional attributes.}
    \label{fig:levelprofile}
\end{figure}

This experiment focuses on profile generation for dimensional attributes.
We refer to a single level of a dimension in the Knowledge Graph including 10k members, and on synthetic
data sources of increasing cardinality, i.e. $\{$10k, 100k, 1M$\}$, including a dimensional attribute with increasing noise, i.e. from 0\% to 50\% in steps of 10\%. Noise refers to the percentage of values in the dimensional attribute that cannot be mapped to any member of the corresponding level.
Values are randomly sampled with replacement from the set of level's members. In Figure \ref{fig:levelprofile}, the average running time (in seconds) and standard deviation are shown varying both the cardinality of the source (expressed in the x-axis, logarithmic scale) and noise (indicated by colors).
For the sake of readability, we report only the times obtained with 0\%, 30\% and 50\% of noise.
We can conclude that profile calculation exhibits an almost linear  growth  with respect to the cardinality of the source. As expected, the time required for computation decreases with increasing noise, as the noise is filtered out during profile generation.

\subsection{Measures and descriptive attributes}
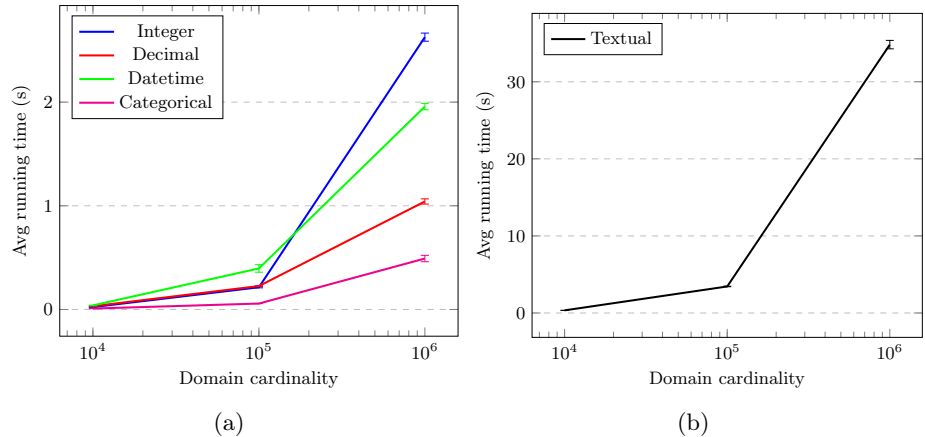
\begin{figure}[htb]
  \begin{subfigure}{0.5\textwidth}
  \resizebox{\textwidth}{!}{
    \begin{tikzpicture}
    \begin{axis}[
        xlabel={Domain cardinality},
        xmode = log,
        ylabel={Avg running time (s)},
        grid style=dashed,
        legend pos= north west,
        xminorticks=true,
        ymajorgrids=true,
    ]
    \addplot[color=blue, line width=1pt, solid, mark=circle, error bars/.cd, y dir=both, y explicit]
    table[x index=0, y index=1, y error index=2, col sep=comma]{mean_integer_time_processing.csv};
    \addlegendentry{Integer}

    \addplot[color=red, line width=1pt, solid, mark=circle, error bars/.cd, y dir=both, y explicit]
    table[x index=0, y index=1, y error index=2, col sep=comma]{mean_real_time_processing.csv};
    \addlegendentry{Decimal}

    \addplot[color=green, line width=1pt, solid, mark=circle, error bars/.cd, y dir=both, y explicit]
    table[x index=0, y index=1, y error index=2, col sep=comma]{mean_date_time_processing.csv};
    \addlegendentry{Datetime}

    \addplot[color=magenta, line width=1pt, solid, mark=circle, error bars/.cd, y dir=both, y explicit]
    table[x index=0, y index=1, y error index=2, col sep=comma]{mean_categorical_time_processing.csv};
    \addlegendentry{Categorical}
    \end{axis}
  \end{tikzpicture}}
  \caption{}
  \label{fig:attr_profile}
  \end{subfigure}
  \begin{subfigure}{0.5\textwidth}
	\centering
  \resizebox{\textwidth}{!}{
 \begin{tikzpicture}
    \begin{axis}[
        xlabel={Domain cardinality},
        xmode = log,
        ylabel={Avg running time (s)},
        grid style=dashed,
        legend pos= north west,
        xminorticks=true,
        ymajorgrids=true,
    ]
    \addplot[color=black, line width=1pt, solid, mark=circle, error bars/.cd, y dir=both, y explicit]
    table[x index=0, y index=1, y error index=2, col sep=comma]{mean_string_time_processing.csv};
    \addlegendentry{Textual}
    \end{axis}
  \end{tikzpicture}}
  \caption{}
  \label{fig:attr_profile_str}
   \end{subfigure}
     \caption{Average running time and standard deviation required for profile calculation of measures or descriptive attributes.}
\end{figure}

This experiment was conducted generating data sources with a fixed number of attributes, one per type, varying the cardinality $card$ in the set $\{$10k, 100k and 1M$\}$. Values are generated according to their type:
\begin{itemize}
    \item for integer and decimal types, values are uniformly distributed in the range [0,card]; for decimal ones, values have a precision of five decimal places;
    \item for categorical types, values are uniformly distributed from a fixed set of four categories;
    \item for datetime types, values are uniformly distributed between a temporal interval of seventy years;
    \item for textual types, values are composed by five words uniformly sampled from a text file containing a total of 83'740 words.
    \end{itemize}
In order to increase the variability, the experiment is repeated ten times and sources are regenerated at each iteration.
In Figure \ref{fig:attr_profile} and in Figure \ref{fig:attr_profile_str}, the average running time (expressed in seconds in the y-axis) and standard deviation are shown varying the cardinality of the source (expressed in the x-axis, logarithmic scale) and the  attribute type (indicated by colors).
We can conclude that execution time 
is almost linear with the cardinality of the source. 
In particular, the execution time for textual attributes is, on average, higher by an order of magnitude if compared to the other types.
This is due to the preprocessing of textual attributes, which includes tokenization of the text and removal of stop-words, following a NLP pipeline.
On the other hand, profile computation for integer attributes is more expensive than other descriptive attributes because of the computation of the data distribution, which in the experiment was based on a fixed-width binning.

In conclusion, the computation times for profile generation appears to be feasible for scenarios involving stable or slowly changing data sources. 
In case of more dynamic scenarios, with continuously updated data sources, strategies for incremental update of profiles are needed. For most of the considered metadata types, the update can be computed straightforwardly, with the exception of the distribution for numeric values. Specifically, if the distribution of newly added data significantly differs from the existing distribution, recalculating or adjusting the profile will be required to accurately reflect the updated data characteristics.

\section{Conclusion}
\label{sec:conclusion}
This paper presented a metadata model for multidimensional sources in the context of data ecosystems, which relies on a graph-based representation and a Knowledge Graph including definitions for dimensions and measures. The model includes metadata for data sources and specific metadata properties for different types of attributes.
Possible future developments include the alignment of properties and classes of the metadata model with existing vocabularies, and the definition of a governance approach  for the management of source metadata, e.g., to support dynamic update of the source profile for incremental sources and efficient storage of metadata, and to address scalability issues. We also aim to exploit the model to support data exploration, analysis and querying in the context of Data Lake(house)s 
 \cite{diamantini2024analytic}.
\subsection*{Acknowledgments}
  Cristina Rossetti has received funding from the MUR – DM 118/2023 as part of the project PNRR-NGEU.
\bibliographystyle{splncs04}
 \bibliography{bibliography}

\begin{thebibliography}{10}
\providecommand{\url}[1]{\texttt{#1}}
\providecommand{\urlprefix}{URL }
\providecommand{\doi}[1]{https://doi.org/#1}

\bibitem{abedjan2014profiling}
Abedjan, Z., Gr{\"u}tze, T., Jentzsch, A., Naumann, F.: Profiling and mining
  rdf data with prolod++. In: 2014 IEEE 30th International Conference on Data
  Engineering. pp. 1198--1201. IEEE (2014)

\bibitem{auer2012lodstats}
Auer, S., Demter, J., Martin, M., Lehmann, J.: Lodstats--an extensible
  framework for high-performance dataset analytics. In: Knowledge Engineering
  and Knowledge Management: 18th International Conference, EKAW 2012, Galway
  City, Ireland, October 8-12, 2012. Proceedings 18. pp. 353--362. Springer
  (2012)

\bibitem{EOSC}
Corcho, O., Eriksson, M., Kurowski, K., et~al.: {EOSC Interoperability
  Framework: Report from the EOSC Executive Board WorkingGroups FAIR and
  Architecture}. \url{doi: 10.2777/620649} (2021)

\bibitem{diamantini2018new}
Diamantini, C., Giudice, P.L., Musarella, L., Potena, D., Storti, E., Ursino,
  D.: A new metadata model to uniformly handle heterogeneous data lake sources.
  In: New Trends in Databases and Information Systems: ADBIS 2018 Short Papers
  and Workshops. pp. 165--177. Springer (2018)

\bibitem{diamantini2023assessment}
Diamantini, C., Mele, A., Potena, D., Storti, E.: Assessment of data quality
  through multi-granularity data profiling. In: European Conference on Advances
  in Databases and Information Systems. pp. 195--209. Springer (2023)

\bibitem{diamantini2024analytic}
Diamantini, C., Potena, D., Storti, E.: Analytic processing in data lakes: A
  semantic query-driven discovery approach. Information Systems Frontiers pp.
  1--19 (2024)

\bibitem{GAIA-X}
GAIA-X: {Gaia-x - Architecture Document 22.04 Release}.
  \url{https://gaia-x.eu/wp-content/uploads/2022/06/Gaia-x-Architecture-Document-22.04-Release.pdf}
  (2022)

\bibitem{DFM98}
Golfarelli, M., Maio, D., Rizzi, S.: {The Dimensional Fact Model: a conceptual
  model for data warehouses }. International Journal of Cooperative Information
  Systems  \textbf{07}(02-03),  215--247 (1998)

\bibitem{hoseini2024survey}
Hoseini, S., Theissen-Lipp, J., Quix, C.: A survey on semantic data management
  as intersection of ontology-based data access, semantic modeling and data
  lakes. Journal of Web Semantics  \textbf{81},  100819 (2024)

\bibitem{langegger2009rdfstats}
Langegger, A., Woss, W.: Rdfstats-an extensible rdf statistics generator and
  library. In: 2009 20th International Workshop on Database and Expert Systems
  Application. pp. 79--83. IEEE (2009)

\bibitem{oram2015managing}
Oram, A.: Managing the Data Lake: Moving to Big Data Analysis. O'Reilly Media
  (2015)

\bibitem{sawadogo2021data}
Sawadogo, P., Darmont, J.: On data lake architectures and metadata management.
  Journal of Intelligent Information Systems  \textbf{56},  97--120 (2021)

\bibitem{sawadogo2019metadata}
Sawadogo, P., Kibata, T., Darmont, J.: Metadata management for textual
  documents in data lakes. arXiv preprint arXiv:1905.04037  (2019)

\bibitem{IDS-RAM4}
Steinbuss, S., Otto, B., Teuscher, A., Pettenpohl, H.: {International Data
  Space Association Reference Architecture Model, v.4.0}.
  \url{https://docs.internationaldataspaces.org/ids-ram-4/} (2022)

\bibitem{wilkinson2016fair}
Wilkinson, M.D., Dumontier, M., Aalbersberg, I.J., Appleton, G., Axton, M.,
  Baak, A., Blomberg, N., Boiten, J.W., da~Silva~Santos, L.B., Bourne, P.E.,
  et~al.: The fair guiding principles for scientific data management and
  stewardship. scientific data 3 (2016). Number  \textbf{1},  160018 (2016)

\end{thebibliography}
\end{document}